\documentclass[amsmath,amssymb,twocolumn,superscriptaddress,floatfix]{revtex4-1}
\usepackage{graphicx}
\usepackage{dcolumn}
\usepackage{bm}
\usepackage{hyperref}
\usepackage{color}
\definecolor{dark-blue}{rgb}{0.0,0.0,0.5}
\definecolor{dark-green}{rgb}{0,0.5,0.0}
\hypersetup{
    colorlinks, linkcolor={dark-blue},
    citecolor={dark-blue}, urlcolor={dark-blue}
}
\usepackage{multirow}
\usepackage[separate-uncertainty=true,multi-part-units=single]{siunitx}
\usepackage{braket}
\DeclareSIUnit[number-unit-product = {}]
\count{c}

\usepackage[usenames,dvipsnames]{xcolor}

\definecolor{ourBlue}{HTML}{0000FF}
\definecolor{ourOrange}{HTML}{FFA400}
\definecolor{ourGreen}{HTML}{008B00}
\definecolor{ourMagenta}{HTML}{FF34A7}
\definecolor{ourPurple}{HTML}{632280}
\definecolor{ourBrown}{HTML}{5E3100}

\newcommand{\ToneSpin}{\mathit{T}_1^\mathrm{Spin}}
\newcommand{\ToneOrbit}{\mathit{T}_1^\mathrm{Orbit}}

\newcommand{\NV}{\mathrm{NV}^-}

\newcommand{\siv}{\mathrm{SiV}^-}
\newcommand{\SiV}{\mathrm{SiV}^-}

\newcommand{\Dthreed}{\mathrm{D}_\mathrm{3d}}

\begin{document}
\title{All-optical initialization, readout, and coherent preparation of single silicon-vacancy spins in diamond}

\author{Lachlan~J.~Rogers}
\affiliation{Institute for Quantum Optics and Center for Integrated Quantum Science and Technology (IQ$\mathrm{^{st}}$), University Ulm, D-89081 Germany}
\author{Kay~D.~Jahnke}
\affiliation{Institute for Quantum Optics and Center for Integrated Quantum Science and Technology (IQ$\mathrm{^{st}}$), University Ulm, D-89081 Germany}
\author{Mathias~H.~Metsch}
\affiliation{Institute for Quantum Optics and Center for Integrated Quantum Science and Technology (IQ$\mathrm{^{st}}$), University Ulm, D-89081 Germany}
\author{Alp~Sipahigil}
\affiliation{Department of Physics,	Harvard University,	17 Oxford Street, Cambridge, MA 02138, USA}
\author{Jan~M.~Binder}
\affiliation{Institute for Quantum Optics and Center for Integrated Quantum Science and Technology (IQ$\mathrm{^{st}}$), University Ulm, D-89081 Germany}
\author{Tokuyuki~Teraji}
\affiliation{National Institute for Materials Science, 1-1 Namiki, Tsukuba,
Ibaraki 305-0044, Japan}
\author{Hitoshi~Sumiya}
\affiliation{Advanced Materials R\,\&\,D Laboratories, Sumitomo Electric Industries Ltd., Itami, Hyogo 664-0016, Japan}
\author{Junichi~Isoya}
\affiliation{Research Center for Knowledge Communities, University of Tsukuba, 1-2 Kasuga, Tsukuba, Ibaraki 305-8550, Japan}
\author{Mikhail~D.~Lukin}
\affiliation{Department of Physics,	Harvard University,	17 Oxford Street, Cambridge, MA 02138, USA}
\author{Philip~Hemmer}
\affiliation{Electrical \& Computer Engineering Department, Texas A\&M University, College Station, TX 77843, USA}
\author{Fedor~Jelezko}
\affiliation{Institute for Quantum Optics and Center for Integrated Quantum Science and Technology (IQ$\mathrm{^{st}}$), University Ulm, D-89081 Germany}





\begin{abstract}
The silicon-vacancy ($\SiV$) color center in diamond has attracted attention due to its unique optical properties. 
It exhibits spectral stability and indistinguishability that facilitate efficient generation of photons capable of demonstrating quantum interference. 
Here we show high fidelity optical initialization and readout of electronic spin in a single $\SiV$ center with a spin relaxation time of $T_1=\SI{2.4\pm0.2}{\milli \second}$.  
Coherent population trapping (CPT) is used to demonstrate coherent preparation of dark superposition states with a spin coherence time of $T_2^\star=\SI{35\pm3}{\nano \second}$.
This is fundamentally limited by orbital relaxation, and an understanding of this process opens the way to extend coherences by engineering interactions with phonons.
These results establish the $\SiV$ center as a solid-state spin-photon interface.

\end{abstract}

\maketitle


%
Coherent quantum systems combining long-lived quantum memory with efficient coupling to optical photons represent a key resource for realization of quantum networks
\cite{kimble2008quantum}. 
%
Color centers in diamond \cite{aharonovich2014diamond} are attractive candidates owing to unique properties of diamond, which include optically transparency and a high lattice quality that allows spin to function as long-lived quantum memory \cite{balasubramanian2009ultralong}.
The negative silicon-vacancy ($\SiV$) defect in diamond \cite{goss1996twelve-line, hepp2014electronic, rogers2014electronic} has exceptional optical properties that facilitate efficient generation of indistinguishable photons from multiple distinct emitters \cite{rogers2014multiple, sipahigil2014indistinguishable}.
%
Here we show optical initialization and readout of electronic spin in a single $\SiV$ center with a spin relaxation time of $T_1=\SI{2.4\pm0.2}{\milli \second}$.  
%
Two-photon resonance \cite{fleischhauer2005electromagnetically} is used to demonstrate coherent preparation of dark superposition states with a spin coherence time of $T_2^\star=\SI{35\pm3}{\nano \second}$.
Finally, we present the first evidence of hyperfine interaction with a ${}^{29}$Si nuclear spin in $\SiV$ which can potentially be used as a memory qubit \cite{togan2011laser}.
%
Electronic spin coherences are shown to be limited by orbital relaxation, and an understanding of this process opens the way to extend coherences by engineering interactions with phonons \cite{burek2013nanomechanical}.

\begin{figure*}
\includegraphics[width=\textwidth]{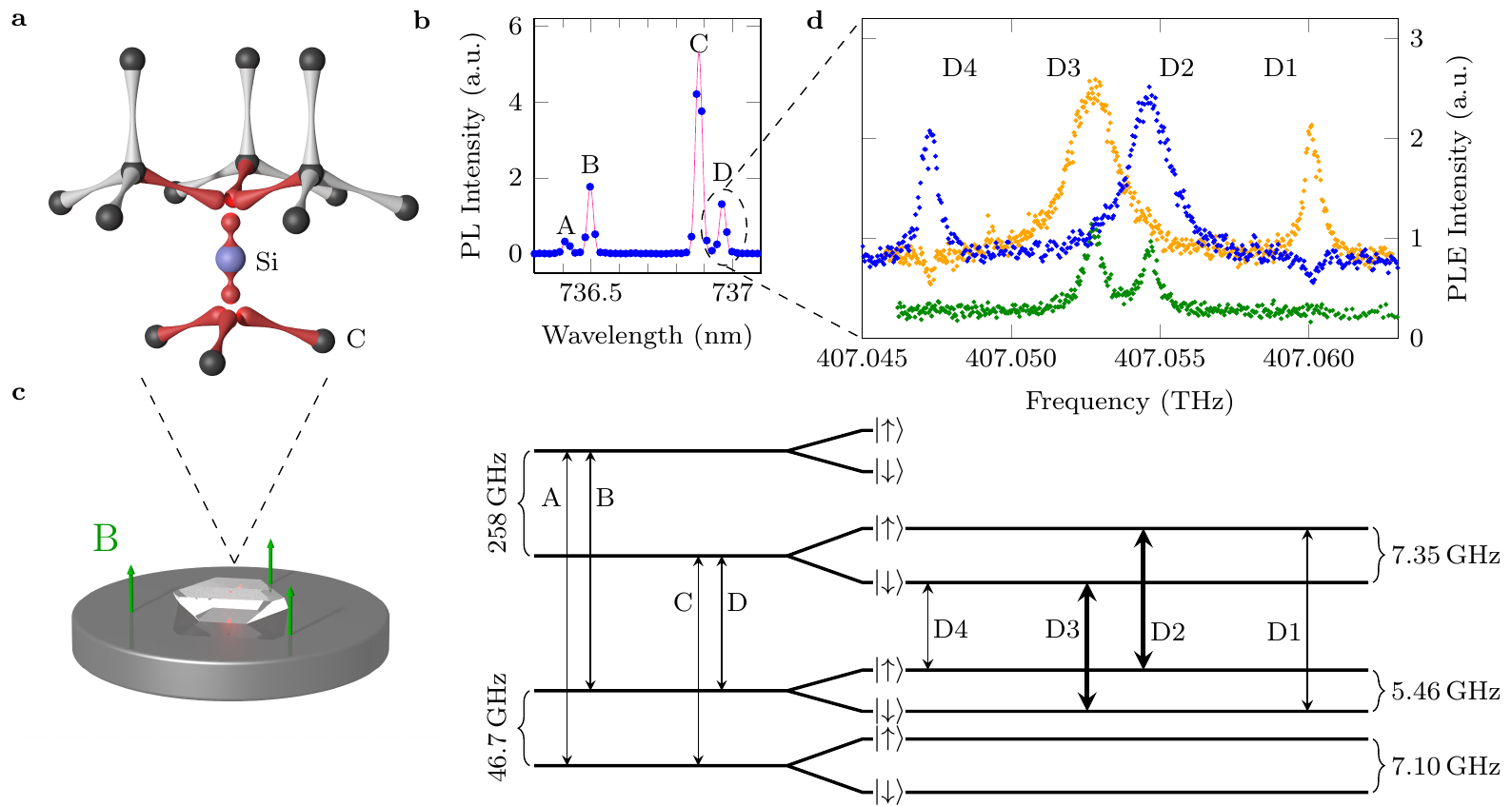}
\caption{
	Optical access to spin levels of $\siv$ in diamond.
	(a) An individual $\SiV$ center consists of a silicon atom between two vacant lattice sites, and is aligned along a $\langle111\rangle$ crystal bond direction. 
	(b) At zero field the ZPL consists of four transitions between spin-orbit branches of the doublet ground and doublet excited states.
	(c) Zeeman splitting of the spin-$\frac{1}{2}$ electronic states ($\ket{\uparrow}$ and $\ket{\downarrow}$ for simplicity) was achieved by mounting the diamond on a neodymium magnet.  
	This produced a $\sim$4.5\,kG field closely aligned with the $\SiV$ centers oriented normal to the \{111\} sample surface.  
	(d) Resonant excitation spectroscopy using a single laser reveals the two spin-conserving transitions D2 and D3 (green curve). 
	Applying a second pump laser on the D2 (orange) or D3 (blue) transition makes the spin-flipping transitions D1 and D4 appear in the excitation spectrum.
	For both cases, the polarity of the D1 and D4 peaks show that these transitions can be utilized to optically pump the spin in the desired state. 
	}
\label{fig:spectra}
\end{figure*}

Quantum information processing efforts in diamond have so far mainly focused on the nitrogen-vacancy ($\NV$) center due to its excellent spin properties at ambient conditions \cite{childress2014atomlike}.
However, small coherent photon generation rates associated with a large phonon sideband and spectral diffusion have limited the development of $\NV$ quantum networks \cite{bernien2013heralded, bernien2012two-photon, sipahigil2012quantum}. 
The ST1 defect \cite{lee2013readout} is the only other color center in diamond that provides optical access to individual spins, but its composition is unknown and its optical spectrum is not ideal. 
The main optical advantage provided by the $\SiV$ center is that its fluorescence is concentrated in a sharp zero-phonon line (ZPL).
The ZPL at 737\,nm contains 70\% of the $\SiV$ emission, and this feature is ideal for single photon source applications \cite{vlasov2009nanodiamond, neu2011single}.
It is spectrally stable and exhibits line widths limited by the excited state lifetime \cite{rogers2014multiple}.
Physically, the $\SiV$ center consists of a single silicon atom replacing two carbon atoms in the diamond lattice, giving the color center $\Dthreed$ symmetry as illustrated in \autoref{fig:spectra}(a) \cite{goss1996twelve-line, rogers2014electronic, hepp2014electronic, dietrich2014isotopically}.  
This geometry makes the $\SiV$ center insensitive to small electric fields \cite{sipahigil2014indistinguishable}, and therefore adds low inhomogeneous broadening to the set of attractive optical properties.
The electronic structure of the $\SiV$ center is reasonably well understood \cite{goss1996twelve-line, hepp2014electronic, rogers2014electronic}, and consists of doubly-degenerate E ground and excited state orbitals which both have electronic spin-$\frac{1}{2}$.
Optical signatures of the electron spin have recently been identified \cite{muller2014optical}, however control of $\SiV$ spin remains an outstanding challenge.

The technique used here to manipulate and measure $\SiV$ spin involves resonant excitation with narrow-band lasers.
At cryogenic temperature the ZPL is split into four transitions (labelled A--D in order of decreasing energy) as shown in \autoref{fig:spectra}(b).
A diamond sample was mounted on a fixed magnet as shown in \autoref{fig:spectra}(c) so that the spin degeneracy was lifted by Zeeman splitting.
Excitation spectra were recorded for a $\SiV$ site aligned to a $\sim$\SI{4.5}{\kilo G} magnetic field (see Methods).
Transition D is the brightest and narrowest line in this configuration \cite{rogers2014electronic, rogers2014multiple}, and was found to split into two sub-peaks as shown in \autoref{fig:spectra}(d).
Two additional peaks were observed when a second laser was pumping either of the two bright transitions.
These four spectral features correspond to the four possible transitions between the Zeeman-split electronic spin sublevels of the orbital branches involved in line D (numbered D1--D4 in order of decreasing energy).
A qualitatively identical pattern of four features was measured for transition C (Supplementary Information), allowing a determination of the Zeeman splittings illustrated in \autoref{fig:spectra}(d).

\begin{figure}
\includegraphics[width=\columnwidth]{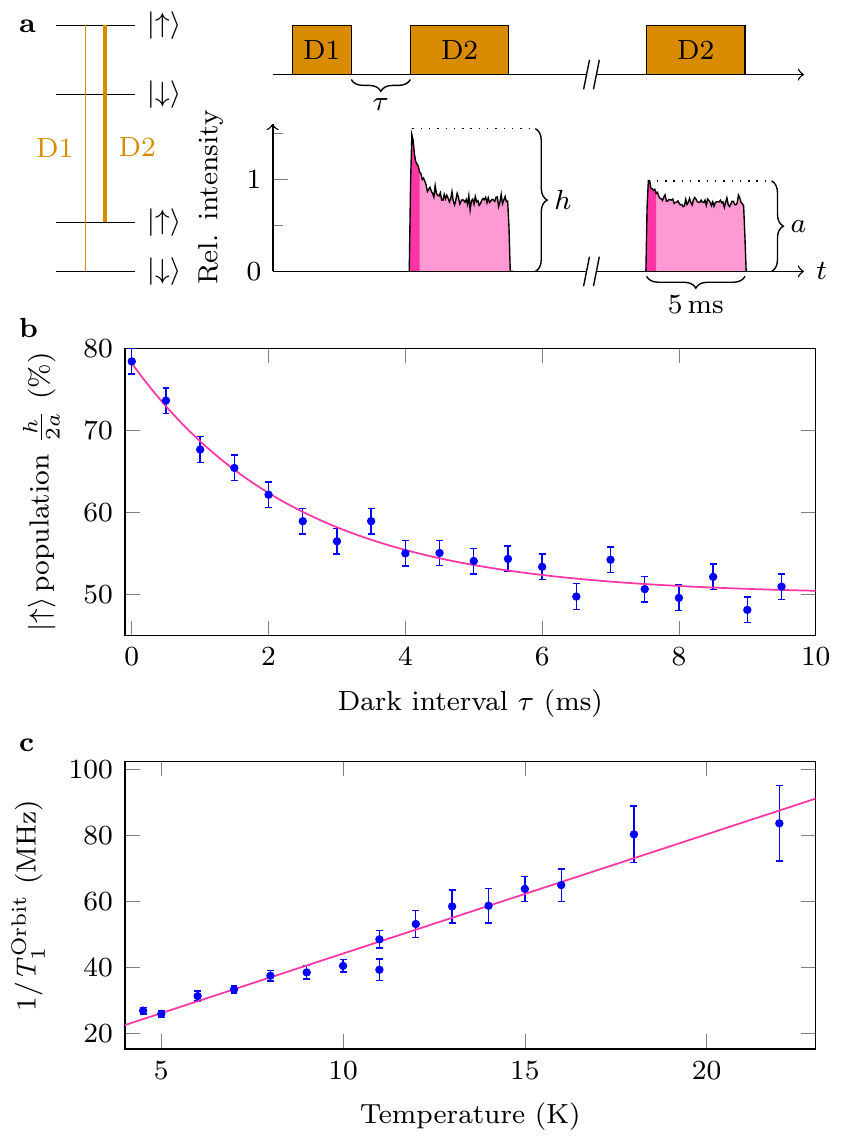}
\caption{
	Optical initialization and readout of $\siv$ spin.
	(a) A laser pulse resonant to transition D1 flips the spin but does not produce measurable fluorescence.
	This spin initialization can be read using a laser pulse on the cycling transition D2.
	It appears as a leading-edge peak of the fluorescence which decays in about \SI{500}{\micro\second}, here shaded darker, which contained on average one photon per pulse.
	After a long dark interval $\tau$ the peak height reduces to an asymptotic value $a$.  
	(b) The reduction of $h$ with increasing $\tau$ gives the spin relaxation time to be $\ToneSpin=\SI{2.4\pm0.2}{\milli \second}$.
	Interpreting the asymptotic limit $a$ to corresopnd to thermal spin population suggests a spin initialization fidelity of $h_{\tau=0}/2a=78\%$.
	(c) Similar pulsed measurements gave the orbital relaxation time at 4.5\,K to be $\ToneOrbit=\SI{38\pm1}{\nano \second}$. 
	The orbital relaxation rate $1/\ToneOrbit$ increased linearly with temperature.
}
\label{fig:relaxation_rates}
\end{figure}

Although optical transitions are spin-conserving, any off-axis magnetic field results in a different quantization axis between the ground and the excited states due to the difference in spin-orbit couplings \cite{hepp2014electronic}. 
In our experiments, the weak off-axis field component introduces a small mixing between spin states making the spin-flipping transitions (D1 and D4) weakly allowed.
These transitions therefore pump the spin into a dark state after a few optical cycles and are not visible in single laser scans.
\autoref{fig:relaxation_rates}(a) shows time-resolved fluorescence measurements where we utilized the D1 transition to optically pump the spin in a dark state and the cycling D2 transition to efficiently readout the resulting spin state. 
The fluorescence produced by the D2 transition had a prominent leading edge when the interpulse delay was small which is evidence of optical pumping causing spin initialization.
The decay of the leading-edge height $h$ with increasing pulse separation $\tau$ establishes the spin relaxation time to be $\ToneSpin=\,$\SI{2.4\pm0.2}{\milli \second}, as shown in \autoref{fig:relaxation_rates}(b)

For a $\siv$ center misaligned about 20 degrees from the magnetic field, the spin relaxation time was measured to be $\ToneSpin =\,\SI{3.4}{\micro\second}$ (Supplementary Information).
When the field was misaligned by $\sim70$ degrees the spin relaxation time reduced to about \SI{60}{\nano\second}.
This sharp decrease in spin relaxation time is a direct result of the spin state mixing induced by the off-axis magnetic field. 
Another effect of this spin mixing is that the peaks D1 and D4 start to appear in excitation spectra without the need for a second repumping laser.
The maximum peak height $h$ in the well-aligned field configuration (\autoref{fig:relaxation_rates}(a)) indicates that we achieved a 78\% spin initialization fidelity.
Given the weak D1 transition dipole for aligned magnetic fields, this incomplete initialization is limited by laser intensity due to competition between the long spin relaxation and slow optical pumping rates on D1.
Equivalent measurements with the field misaligned by about 20 degrees, where the spin-flipping transition is stronger, demonstrated an initialization fidelity of at least 95\% (Supplementary Information).

These observations indicate that the $\SiV$ center provides robust access to both spin-flipping lambda transitions as well as long-cycling spin-conserving transitions which enable optical readout of the spin state.
During each D2 laser pulse shown in \autoref{fig:relaxation_rates}, the fluorescence decayed from its initial height $h$ in about \SI{500}{\micro \second}.  
This shaded readout interval is long given that the D2 transition was excited above saturation, indicating the cyclicity of this transition.
Such long-cycling transitions should allow single-shot spin readout with high fidelity depending primarily on photon collection efficiency \cite{robledo2011high-fidelity}.
From the steady-state photon count rate observed here we estimate the readout peak contained on average one photon.

The $\SiV$ ground states  also have an orbital degeneracy which can introduce additional dynamics beyond spin $\ToneSpin$ processes \cite{stoneham2001theory,ham1965dynamical}. 
A sequence of resonant excitation pulses was used to investigate the orbital relaxation time between the ground state spin-orbit branches (see Methods), and it was found to be only $\ToneOrbit=\SI{38\pm1}{\nano \second}$ at 4.5\,K.
This is much shorter than $\ToneSpin$, which indicates that the orbital relaxation must be highly spin-conserving as expected for electron-phonon interactions.
Measurements of $\ToneOrbit$ were repeated for temperatures up to 22\,K, and the rate was found to vary linearly with temperature as shown in \autoref{fig:relaxation_rates}(c).
This is evidence of a direct single-phonon mechanism for this fast relaxation process \cite{jahnke2014phonon}, which has significant implications for the spin coherence time discussed below.


\begin{figure*}
\includegraphics[width=\textwidth]{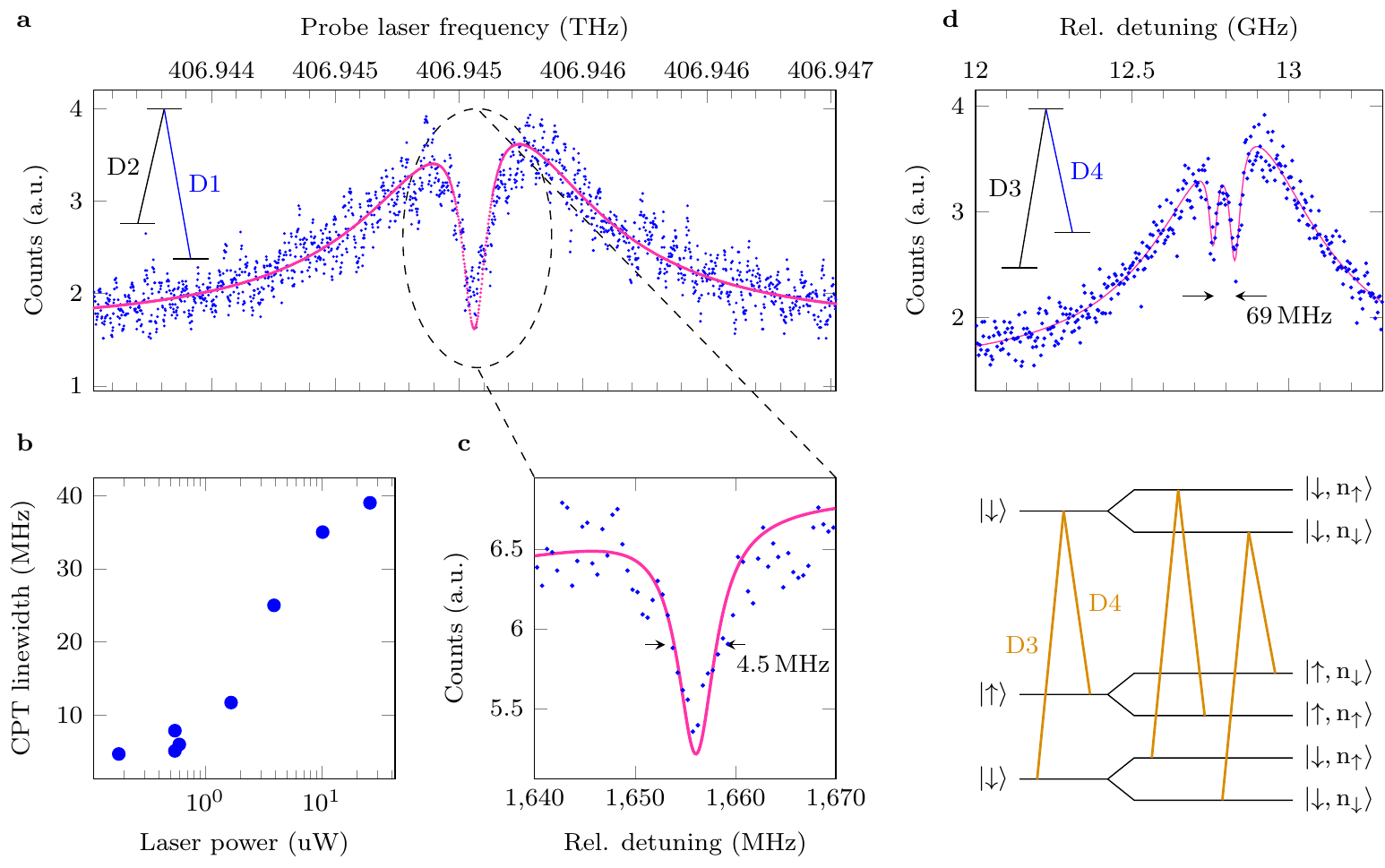}
\caption{
	Preparation of coherent dark states.
	(a) Typical excitation spectrum for a probe laser scanned across transition D1 while a second pump laser was applied resonant to D2. 
	These transitions form a $\Lambda$-scheme giving rise to the coherent population trapping (CPT) phenomenon, producing the characteristic sharp dip in the spectrum.
	(b) By reducing the incident laser powers it was possible to avoid power-broadening of the CPT dip.
	(c) The narrowest dip measured had a FWHM of \SI{4.5\pm0.3}{\mega \hertz}, corresponding to a spin coherence time of $T_2^\star=\SI{35\pm3}{\nano\second}$.
	(d) A double CPT dip was observed for $\SiV$ centers containing the isotope ${}^{29}\mathrm{Si}$ which has nuclear spin $I=\frac{1}{2}$.
	The nuclear Zeeman shift is reversed for the opposite electronic spin, and so the two nuclear-spin-conserving $\Lambda$-schemes occur at different relative laser detunings.		
	The splitting of 69\,MHz between the CPT dips suggests a nuclear hyperfine splitting on the order of 35\,MHz.
}
\label{fig:eit}
\end{figure*}

With this understanding of the transitions and the population dynamics, it is possible to identify $\Lambda$-schemes that allow for the coherent manipulation of spin.  
In general, a $\Lambda$-scheme consisting of two allowed transitions from different ground states to a common excited state provides the ability to optically prepare a dark superposition of the ground states \cite{fleischhauer2005electromagnetically}. 
This phenomenon is known as coherent population trapping (CPT) and can be used to coherently control ground state coherences.

To observe CPT a pump laser was tuned to transition D3 and a probe laser was scanned across transition D4 (see Methods).
The excitation peak for transition D4 was observed to contain a sharp two-photon resonance dip as shown in \autoref{fig:eit}(a).
The width of this CPT dip was measured for various laser powers (\autoref{fig:eit}(b)), and the limiting width in the absence of power broadening was found to be about \SI{4.5\pm0.3}{\mega \hertz}  as shown in \autoref{fig:eit}(c) .
This is far below the transform-limited optical linewidth of 94\,MHz \cite{rogers2014multiple}, which confirms CPT and corresponds to a spin coherence time of $T_2^\star=\SI{35\pm3}{\nano\second}$.
This coherence time is similar to $\ToneOrbit$, indicating that the rapid switching between orbital branches leads to dephasing of the electronic spin even while its polarization is maintained. 
While this dramatically limits the usefulness of electronic spin in $\siv$, our identification of the orbital switching mechanism makes it possible to consider techniques to overcome this limit.

One approach to create a long-lived memory is provided by $\siv$ centers containing the  ${}^{29}\mathrm{Si}$ isotope, which has nuclear spin.
It is possible to identify such $\siv$ centers due to an isotopic shift of the ZPL \cite{dietrich2014isotopically}.
We observed that CPT for ${}^{29}\mathrm{Si}$ $\siv$ sites produces a double two-photon resonance condition as shown in \autoref{fig:eit}(c).
This doublet arises due to the hyperfine interaction with the ${}^{29}\mathrm{Si}$ nuclear spin ($I=1/2$) which gives rise to a two-photon resonance condition at two different detunings as shown in \autoref{fig:eit}(d).
This observation of  ${}^{29}\mathrm{Si}$ hyperfine splitting in $\siv$ raises the possibility of optical access to nuclear spin, which should have much longer coherence times than electron spins \cite{childress2006coherent, jacques2009dynamic,maurer2012room-temperature}.

There are a number of other approaches that could be used to improve the coherence time of the electronic spin in $\siv$.
Since $T_2^\star$ is limited by orbital relaxation that arises from single-phonon processes, most of these ideas revolve around limiting the availability of phonons at the relevant frequency ($\sim47$\,GHz).
Cooling below 1\,K will freeze out the upward exchange process between the ground state spin-orbit branches.  
Using the lower branch (i.e. transition C) would then allow electronic spin to be coherently manipulated without dephasing.
Alternatively, one can reduce the phonon density of states at this frequency. 
This could be realized by using nanodiamonds smaller than the phonon wavelength ($\sim$250\,nm) which cannot support low frequency vibrations or using fabricated structures to realize a phonon band gap \cite{burek2013nanomechanical} around ($\sim47$\,GHz). 
We expect both approaches to reduce the orbital relaxation rate and therefore increase the spin coherence time.


Our experiments demonstrate that resonant excitation can be used to initialize and control electronic spin in individual $\SiV$ centers.
The electron spin relaxation time is $\ToneSpin=\SI{2.4\pm0.2}{\milli\second}$ in a well aligned magnetic field, and extremely long-cycling optical transitions make it possible to efficiently read out the spin state.
Coherent population trapping demonstrated the preparation of dark spin-superposition states with a coherence time of $T_2^\star=\SI{35\pm3}{\nano\second}$.
We have shown this to be limited by a fast orbital relaxation process, and proposed various techniques for extending the coherence time.
Hyperfine splitting of ${}^{29}\mathrm{Si}$ nuclear spin was observed in $\SiV$ for the first time.
These results open the door for exploration of long-lived quantum memories based on $\SiV$ centers and the development of spin-photon interfaces benefiting from unique optical properties.

{\em Note}: During the preparation of this manuscript we became aware of complementary work reporting coherent population trapping with $\SiV$ centers \cite{pingault2014all-optical}.


\section*{Methods}

\subsubsection*{Sample preparation}

Two diamond samples were used in this study, the first to provide a \{111\} crystal face and the second to allow ${}^{29}\mathrm{Si}$-containing $\SiV$ sites to be found.

A high-quality and high-purity single-crystal diamond (about 6--7\,mm in
diameter) containing very few crystal defects and few nitrogen
impurities ($\sim2$\,ppb) was synthesized by the temperature gradient method
at a high pressure and high temperature (HPHT) condition of 5.5 GPa and \SI{1350}{\celsius}
under highly precise temperature control \cite{sumiya2012large}. 
A high purity Fe–Co alloy was used as the solvent, and Ti was added to the solvent as a nitrogen getter.
High-crystalline-quality diamond crystals were used as seed crystals.
The grown crystal was sliced with a laser parallel to the \{111\} surface to produce a plate.
In this study, we observed the $\SiV$ centers in the as-grown \{111\} surface.

The density of $\SiV$ sites in this \{111\} plate was too low to find ${}^{29}\mathrm{Si}$, which has a natural abundance of only 4.7\%.  
To find silicon nuclear spin, a second sample was used which had been grown using a microwave plasma-assisted chemical-vapor-deposition (MPCVD) technique.
The fabrication details of this sample have been reported previously \cite{rogers2014multiple}.
Silicon was etched by the plasma from a silicon-carbide plate, ensuring the incorporation of single $\SiV$ centers at a suitable concentration.  
This incorporation also ensured ${}^{29}\mathrm{Si}$ was present in its natural isotopic abundance.

\subsubsection*{Resonant excitation spectroscopy}

The diamond sample was cooled to approximately 5\,K in a continuous flow cryostat, and imaged using a home-built confocal fluorescence microscope.
Resonant excitation spectra were measured by scanning a high-resolution (100\,kHz) laser across the frequency of a $\SiV$ transition and recording fluorescence.
Laser scatter was excluded from the detection path using a 750--810\,nm band pass filter, which transmitted fluorescence from the $\siv$ sideband.
Since the majority of $\SiV$ fluorescence is contained in the ZPL, this arrangement significantly reduces the overall photon collection efficiency.

For the experiments in \autoref{fig:eit}(b-c), the two excitation frequencies were generated using a single laser which was modulated using a high bandwidth electro-optic amplitude modulator operated in the linear response regime. 
The modulation frequency was scanned across the two-photon detuning using microwave sources, and
the CPT linewidth measurements are therefore insensitive to laser frequency noise. 
Probe field intensity was varied by changing modulation amplitude.
Laser power limitations with this equipment made it difficult to address the dark transitions for the well-aligned field.
Measurements were made on a second $\SiV$ site oriented at about 70 degrees to the field, accounting for the lower D1--D2 detuning.

In order to more clearly resolve the nuclear splitting in \autoref{fig:eit}(d), a stronger magnetic field was applied.  
This was achieved by mounting the sample on a pattern of four cubic magnets.
This accounts for the increased relative detuning between transitions D3 and D4

\subsubsection*{Spin relaxation time}

As described in the text and illustrated in \autoref{fig:relaxation_rates}(a), laser pulses on transitions D1 and D2 were used to measure the spin relaxation time.
These pulses were produced using acousto-optical modulators, and the laser output was measured to be constant over the duration of the pulse.
Laser pulses produced in this manner had a rise/fall time of about 60\,ns, and the extinction ratio was measured at the confocal microscope to be about 60\,dB.
Time-resolved photon counts were recorded using a FAST ComTec MCS26A data acquisition card configured to have \SI{6.4}{\micro\second} time bins.

\subsubsection*{Orbital relaxation time}

In order to isolate the orbital relaxation process from any spin relaxation processes we performed this measurement at zero magnetic field. 
Since the time scale was found to be fast compared to the acousto-optical modulator switching duration, 
the acousto-optical modulator was replaced by an electro-optical modulator capable of much faster switching. 
The laser pulses were measured to have rise/fall times of 1\,ns.
Pulses of 80\,ns were applied resonant to transition D, and photon counts were recorded at the maximum timing resolution of 200\,ps.
There was a leading-edge peak on the fluorescence pulse, and the decay from the initial intensity corresponds to population being pumped to the other ground state branch (decaying via transition C).
The time between subsequent laser pulses was varied, and the recovery of the leading-edge peak gave the orbital relaxation time. 
A detailed experimental and theoretical study of the orbital relaxation processes within the excited and ground states between 4-350 K is presented in Ref \cite{jahnke2014phonon}.

\section*{References}
\bibliographystyle{naturemag}
\bibliography{paper_SiV_eit}

\begin{thebibliography}{10}
\expandafter\ifx\csname url\endcsname\relax
  \def\url#1{\texttt{#1}}\fi
\expandafter\ifx\csname urlprefix\endcsname\relax\def\urlprefix{URL }\fi
\providecommand{\bibinfo}[2]{#2}
\providecommand{\eprint}[2][]{\url{#2}}

\bibitem{kimble2008quantum}
\bibinfo{author}{Kimble, H.~J.}
\newblock \bibinfo{title}{The quantum internet}.
\newblock \emph{\bibinfo{journal}{Nature}} \textbf{\bibinfo{volume}{453}},
  \bibinfo{pages}{1023--1030} (\bibinfo{year}{2008}).

\bibitem{aharonovich2014diamond}
\bibinfo{author}{Aharonovich, I.} \& \bibinfo{author}{Neu, E.}
\newblock \bibinfo{title}{Diamond nanophotonics}.
\newblock \emph{\bibinfo{journal}{Advanced Optical Materials}}
  \bibinfo{pages}{Early View, doi: 10.1002/adom.201400189}
  (\bibinfo{year}{2014}).

\bibitem{balasubramanian2009ultralong}
\bibinfo{author}{Balasubramanian, G.} \emph{et~al.}
\newblock \bibinfo{title}{Ultralong spin coherence time in isotopically
  engineered diamond}.
\newblock \emph{\bibinfo{journal}{Nature Materials}}
  \textbf{\bibinfo{volume}{8}}, \bibinfo{pages}{383--387}
  (\bibinfo{year}{2009}).

\bibitem{goss1996twelve-line}
\bibinfo{author}{Goss, J.~P.}, \bibinfo{author}{Jones, R.},
  \bibinfo{author}{Breuer, S.~J.}, \bibinfo{author}{Briddon, P.~R.} \&
  \bibinfo{author}{{\"O}berg, S.}
\newblock \bibinfo{title}{The twelve-line 1.682 {eV} luminescence center in
  diamond and the vacancy-silicon complex}.
\newblock \emph{\bibinfo{journal}{Physical Review Letters}}
  \textbf{\bibinfo{volume}{77}}, \bibinfo{pages}{3041--3044}
  (\bibinfo{year}{1996}).

\bibitem{hepp2014electronic}
\bibinfo{author}{Hepp, C.} \emph{et~al.}
\newblock \bibinfo{title}{Electronic structure of the silicon vacancy color
  center in diamond}.
\newblock \emph{\bibinfo{journal}{Physical Review Letters}}
  \textbf{\bibinfo{volume}{112}}, \bibinfo{pages}{036405}
  (\bibinfo{year}{2014}).

\bibitem{rogers2014electronic}
\bibinfo{author}{Rogers, L.~J.} \emph{et~al.}
\newblock \bibinfo{title}{Electronic structure of the negatively charged
  silicon-vacancy center in diamond}.
\newblock \emph{\bibinfo{journal}{Physical Review B}}
  \textbf{\bibinfo{volume}{89}}, \bibinfo{pages}{235101}
  (\bibinfo{year}{2014}).

\bibitem{rogers2014multiple}
\bibinfo{author}{Rogers, L.~J.} \emph{et~al.}
\newblock \bibinfo{title}{Multiple intrinsically identical single-photon
  emitters in the solid state}.
\newblock \emph{\bibinfo{journal}{Nature Communications}}
  \textbf{\bibinfo{volume}{5}}, \bibinfo{pages}{4739} (\bibinfo{year}{2014}).

\bibitem{sipahigil2014indistinguishable}
\bibinfo{author}{Sipahigil, A.} \emph{et~al.}
\newblock \bibinfo{title}{Indistinguishable photons from separated
  silicon-vacancy centers in diamond}.
\newblock \emph{\bibinfo{journal}{Physical Review Letters}}
  \textbf{\bibinfo{volume}{113}}, \bibinfo{pages}{113602}
  (\bibinfo{year}{2014}).

\bibitem{fleischhauer2005electromagnetically}
\bibinfo{author}{Fleischhauer, M.}, \bibinfo{author}{Imamoglu, A.} \&
  \bibinfo{author}{Marangos, J.~P.}
\newblock \bibinfo{title}{Electromagnetically induced transparency: Optics in
  coherent media}.
\newblock \emph{\bibinfo{journal}{Reviews of Modern Physics}}
  \textbf{\bibinfo{volume}{77}}, \bibinfo{pages}{633--673}
  (\bibinfo{year}{2005}).

\bibitem{togan2011laser}
\bibinfo{author}{Togan, E.}, \bibinfo{author}{Chu, Y.},
  \bibinfo{author}{Imamoglu, A.} \& \bibinfo{author}{Lukin, M.~D.}
\newblock \bibinfo{title}{Laser cooling and real-time measurement of the
  nuclear spin environment of a solid-state qubit}.
\newblock \emph{\bibinfo{journal}{Nature}} \textbf{\bibinfo{volume}{478}},
  \bibinfo{pages}{497} (\bibinfo{year}{2011}).

\bibitem{burek2013nanomechanical}
\bibinfo{author}{Burek, M.~J.}, \bibinfo{author}{Ramos, D.},
  \bibinfo{author}{Patel, P.}, \bibinfo{author}{Frank, I.~W.} \&
  \bibinfo{author}{Lon{\v c}ar, M.}
\newblock \bibinfo{title}{Nanomechanical resonant structures in single-crystal
  diamond}.
\newblock \emph{\bibinfo{journal}{Applied Physics Letters}}
  \textbf{\bibinfo{volume}{103}}, \bibinfo{pages}{131904}
  (\bibinfo{year}{2013}).

\bibitem{childress2014atomlike}
\bibinfo{author}{Childress, L.}, \bibinfo{author}{Walsworth, R.} \&
  \bibinfo{author}{Lukin, M.}
\newblock \bibinfo{title}{Atom-like crystal defects: From quantum computers to
  biological sensors}.
\newblock \emph{\bibinfo{journal}{Physics Today}}
  \textbf{\bibinfo{volume}{67}}, \bibinfo{pages}{38--43}
  (\bibinfo{year}{2014}).

\bibitem{bernien2013heralded}
\bibinfo{author}{Bernien, H.} \emph{et~al.}
\newblock \bibinfo{title}{Heralded entanglement between solid-state qubits
  separated by three metres}.
\newblock \emph{\bibinfo{journal}{Nature}} \textbf{\bibinfo{volume}{497}},
  \bibinfo{pages}{86--90} (\bibinfo{year}{2013}).

\bibitem{bernien2012two-photon}
\bibinfo{author}{Bernien, H.} \emph{et~al.}
\newblock \bibinfo{title}{Two-photon quantum interference from separate
  nitrogen vacancy centers in diamond}.
\newblock \emph{\bibinfo{journal}{Physical Review Letters}}
  \textbf{\bibinfo{volume}{108}}, \bibinfo{pages}{043604}
  (\bibinfo{year}{2012}).

\bibitem{sipahigil2012quantum}
\bibinfo{author}{Sipahigil, A.} \emph{et~al.}
\newblock \bibinfo{title}{Quantum interference of single photons from remote
  nitrogen-vacancy centers in diamond}.
\newblock \emph{\bibinfo{journal}{Physical Review Letters}}
  \textbf{\bibinfo{volume}{108}}, \bibinfo{pages}{143601}
  (\bibinfo{year}{2012}).

\bibitem{lee2013readout}
\bibinfo{author}{Lee, S.-Y.} \emph{et~al.}
\newblock \bibinfo{title}{Readout and control of a single nuclear spin with a
  metastable electron spin ancilla}.
\newblock \emph{\bibinfo{journal}{Nature Nanotechnology}}
  \textbf{\bibinfo{volume}{8}}, \bibinfo{pages}{487--492}
  (\bibinfo{year}{2013}).

\bibitem{vlasov2009nanodiamond}
\bibinfo{author}{Vlasov, I.~I.} \emph{et~al.}
\newblock \bibinfo{title}{Nanodiamond photoemitters based on strong narrow-band
  luminescence from silicon-vacancy defects}.
\newblock \emph{\bibinfo{journal}{Advanced Materials}}
  \textbf{\bibinfo{volume}{21}}, \bibinfo{pages}{808--812}
  (\bibinfo{year}{2009}).

\bibitem{neu2011single}
\bibinfo{author}{Neu, E.} \emph{et~al.}
\newblock \bibinfo{title}{Single photon emission from silicon-vacancy colour
  centres in chemical vapour deposition nano-diamonds on iridium}.
\newblock \emph{\bibinfo{journal}{New Journal of Physics}}
  \textbf{\bibinfo{volume}{13}}, \bibinfo{pages}{025012}
  (\bibinfo{year}{2011}).

\bibitem{dietrich2014isotopically}
\bibinfo{author}{Dietrich, A.} \emph{et~al.}
\newblock \bibinfo{title}{Isotopically varying spectral features of silicon
  vacancy in diamond}  (\bibinfo{year}{2014}).
\newblock \bibinfo{note}{Http://arxiv.org/abs/1407.7137}.

\bibitem{muller2014optical}
\bibinfo{author}{M{\"u}ller, T.} \emph{et~al.}
\newblock \bibinfo{title}{Optical signatures of silicon-vacancy spins in
  diamond}.
\newblock \emph{\bibinfo{journal}{Nature Communications}}
  \textbf{\bibinfo{volume}{5}} (\bibinfo{year}{2014}).

\bibitem{robledo2011high-fidelity}
\bibinfo{author}{Robledo, L.} \emph{et~al.}
\newblock \bibinfo{title}{High-fidelity projective read-out of a solid-state
  spin quantum register}.
\newblock \emph{\bibinfo{journal}{Nature}} \textbf{\bibinfo{volume}{477}},
  \bibinfo{pages}{574--578} (\bibinfo{year}{2011}).

\bibitem{stoneham2001theory}
\bibinfo{author}{Stoneham, A.~M.}
\newblock \emph{\bibinfo{title}{Theory of Defects in Solids: Electronic
  Structure of Defects in Insulators and Semiconductors}}
  (\bibinfo{publisher}{Oxford University Press}, \bibinfo{year}{2001}).

\bibitem{ham1965dynamical}
\bibinfo{author}{Ham, F.~S.}
\newblock \bibinfo{title}{Dynamical jahn-teller effect in paramagnetic
  resonance spectra: Orbital reduction factors and partial quenching of
  spin-orbit interaction}.
\newblock \emph{\bibinfo{journal}{Phys. Rev.}} \textbf{\bibinfo{volume}{138}},
  \bibinfo{pages}{A1727--A1740} (\bibinfo{year}{1965}).

\bibitem{jahnke2014phonon}
\bibinfo{author}{Jahnke, K.~D.} \emph{et~al.}
\newblock \bibinfo{title}{Phonon processes for the silicon-vacancy center in
  diamond}.
\newblock \bibinfo{type}{Manuscript} \bibinfo{number}{in preparation}
  (\bibinfo{year}{2014}).

\bibitem{childress2006coherent}
\bibinfo{author}{Childress, L.} \emph{et~al.}
\newblock \bibinfo{title}{Coherent dynamics of coupled electron and nuclear
  spin qubits in diamond}.
\newblock \emph{\bibinfo{journal}{Science}} \textbf{\bibinfo{volume}{314}},
  \bibinfo{pages}{281} (\bibinfo{year}{2006}).

\bibitem{jacques2009dynamic}
\bibinfo{author}{Jacques, V.} \emph{et~al.}
\newblock \bibinfo{title}{Dynamic polarization of single nuclear spins by
  optical pumping of nitrogen-vacancy color centers in diamond at room
  temperature}.
\newblock \emph{\bibinfo{journal}{Physical Review Letters}}
  \textbf{\bibinfo{volume}{102}}, \bibinfo{pages}{057403}
  (\bibinfo{year}{2009}).

\bibitem{maurer2012room-temperature}
\bibinfo{author}{Maurer, P.~C.} \emph{et~al.}
\newblock \bibinfo{title}{Room-temperature quantum bit memory exceeding one
  second}.
\newblock \emph{\bibinfo{journal}{Science}} \textbf{\bibinfo{volume}{336}},
  \bibinfo{pages}{1283--1286} (\bibinfo{year}{2012}).

\bibitem{pingault2014all-optical}
\bibinfo{author}{Pingault, B.} \emph{et~al.}
\newblock \bibinfo{title}{All-optical formation of coherent dark states of
  silicon-vacancy spins in diamond}  (\bibinfo{year}{2014}).
\newblock \bibinfo{note}{Http://arxiv.org/abs/1409.4069}.

\bibitem{sumiya2012large}
\bibinfo{author}{Sumiya, H.} \& \bibinfo{author}{Tamasaku, K.}
\newblock \bibinfo{title}{Large defect-free synthetic type {IIa} diamond
  crystals synthesized via high pressure and high temperature}.
\newblock \emph{\bibinfo{journal}{Japanese Journal of Applied Physics}}
  \textbf{\bibinfo{volume}{51}}, \bibinfo{pages}{090102}
  (\bibinfo{year}{2012}).

\end{thebibliography}

\section*{Acknowledgements}

We acknowledge ERC, EU projects (SIQS, DIADEMS, EQUAM), DFG (FOR 1482, FOR 1493
and SFBTR 21), JST, JSPS KAKENHI (No.\,26246001), BMBF, USARL/ORISE, DARPA QuASAR, NSF ECCS-1202258, and the Sino-German and
Volkswagen foundations for funding. 
%

\section*{Author Contributions}

LR, MM, KJ, JB, and AS performed the experiments, which were conceived by PH, FJ, and ML.
TT, HS, and JI synthesized the samples.
LR, KJ, and MM wrote the manuscript with input from all the authors.

\newpage

\section*{Supplementary Information}

\subsubsection*{Excitation spectra for transitions C1--C4}

\begin{figure}[h]
\includegraphics[width=\columnwidth]{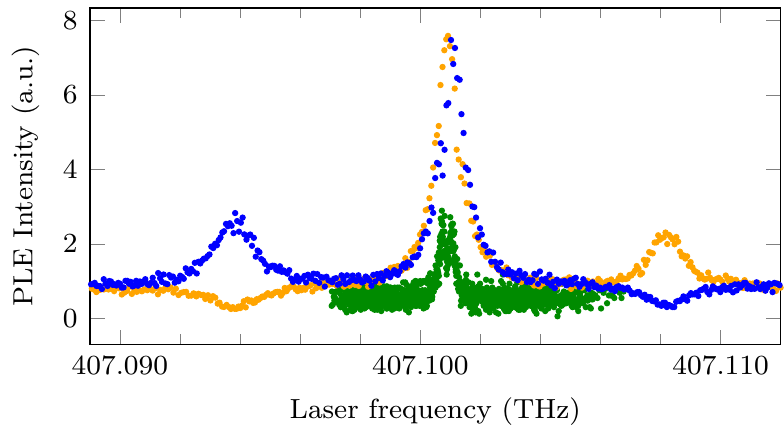}
\caption{ 
	Resonant excitation spectrum of the transitions in line C for the same $\sim$4.5\,kG magnetic field as in \autoref{fig:spectra}.
	As for line D, only the two spin-conserving transitions C2 and C3 (green curve) are visible when scanning with a single laser.
	Applying a second pump laser to C2 (orange) or C3 (blue) makes the spin-flipping transitions C1 and C4 appear in the excitation spectrum.
}
\label{fig:Spec_line_C}
\end{figure}

A single resonant laser was scanned across transition C in the same experimental configuration used for the spectra of transition D in \autoref{fig:spectra}(d).
Two peaks were visible, although they were not clearly resolved.  
This indicates that the spin conserving transitions C2 and C3 are nearly degenerate, and therefore the Zeeman splittings of the ground and excited states are nearly identical (see \autoref{fig:spectra}). 
This makes it more difficult to use transition C for any resonant excitation experiments which rely on exciting C2 or C3 independently.
Again a second pump laser on the C2 or C3 transition made extra peaks appear in the spectrum as shown in \autoref{fig:Spec_line_C} corresponding to the spin flipping transitions C1 and C4.

\subsubsection*{Rate equation modelling of excitation spectra}

\begin{figure}
\includegraphics[width=\columnwidth]{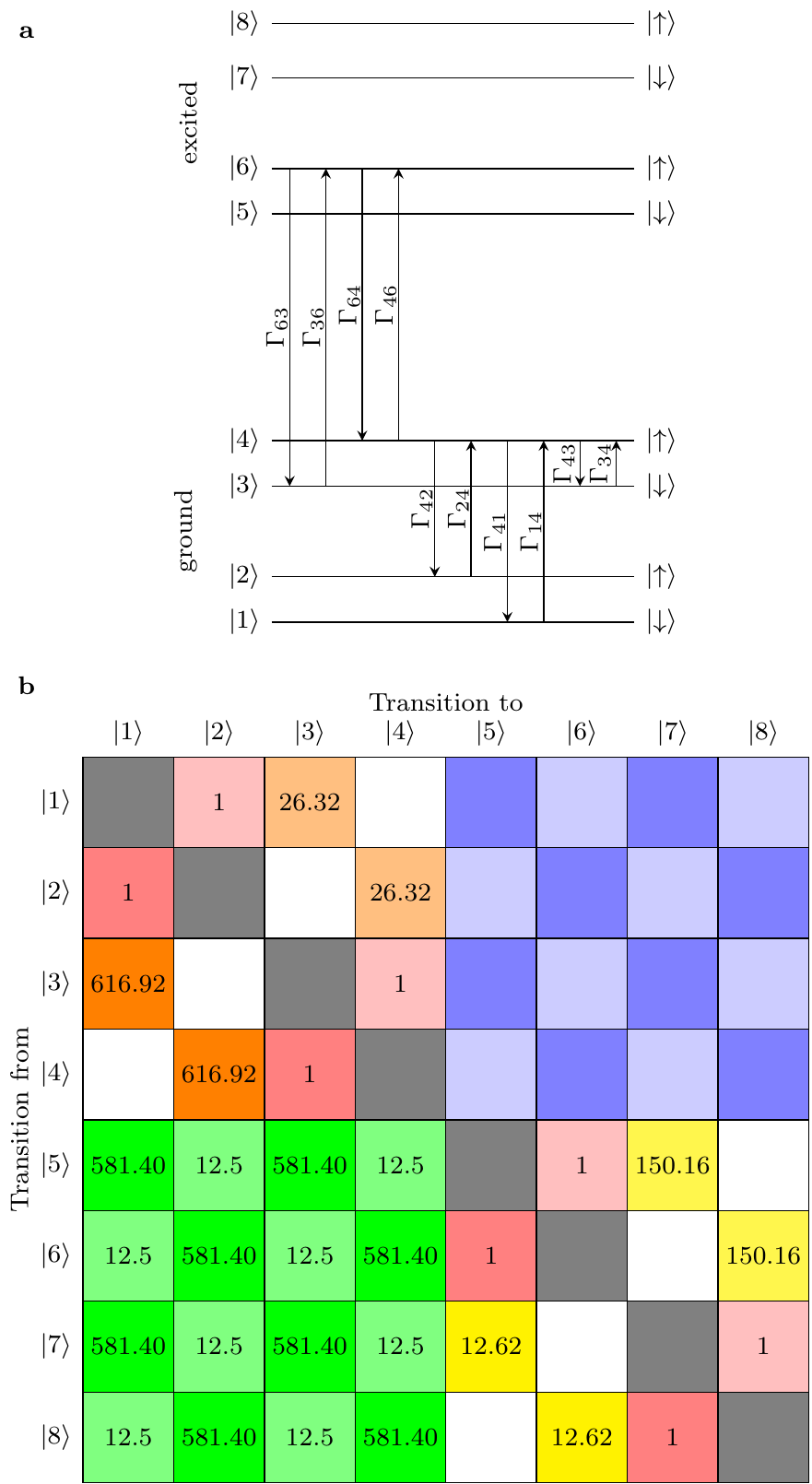}
\caption{
	Rate equation model for simulating excitation spectra.
	(a) The levels are numbered from lowest energy to highest and correspond to the level scheme in \autoref{fig:spectra}.
	For each transition the rate $\Gamma_{ij}$ from level $\ket{i}$ to $\ket{j}$ is not necessarily the same as the rate $\Gamma_{ji}$.
	(b) The matrix describes all the rates used to model the system (in MHz).
	Optical decay transitions are shaded in green with pale colour for spin flipping transitions and dark for spin conserving.
	Optical excitation rates (blue) are derived from the corresponding decay rate, but depend on laser intensity which varies for each transition according to the chosen laser frequency.
	Magenta and pink indicate spin flip events within an orbital branch ($\ToneSpin$), and orange (yellow) represents the orbital mixing in the ground (excited) state.
	For these rates the paler color denotes transition to a higher level while darker indicates decay to a lower level.
}
\label{fig:rates}
\end{figure}

At first it was difficult to interpret the ``dips'' that occur in the resonant excitation spectra for transitions D1, D4, C1, and C4 depending on which spin-conserving transition is pumped (see \autoref{fig:spectra} and \autoref{fig:Spec_line_C}).
A rate-equation simulation was used to determine that such features naturally arise from the spin population dynamics in the Zeeman-split $\SiV$ system.
The model consists of differential equations describing the change in population of each state in terms of the rates in and the rates out.
All eight states were considered, corresponding to the Zeeman-split situation as illustrated in \autoref{fig:rates}(a).
The states were numbered with $\ket{1}$--$\ket{4}$ corresponding to the ground states and $\ket{5}$--$\ket{8}$ comprising the excited state.

The rates for each possible transition are presented in \autoref{fig:rates}(b). 
Rates were obtained from measured results where possible.
The spin-flip rate for a misaligned field and was measured to be \SI{1}{\micro\second} while the radiative lifetime is known to be 1.72\,ns \cite{rogers2014multiple}. 
The orbital mixing rates for the ground states was determined from our orbital relaxation measurement.
The equivalent mixing rates for the exited states were derived from the observed optical line widths \cite{rogers2014multiple} and a Boltzmann factor was applied to impede the upward rates.
The optical excitation rates (shaded blue in \autoref{fig:rates}(b)) were determined for each transition using a lorentzian lineshape to scale the effective laser intensity as a function of detuning.

\begin{figure}[]
\includegraphics[width=\columnwidth]{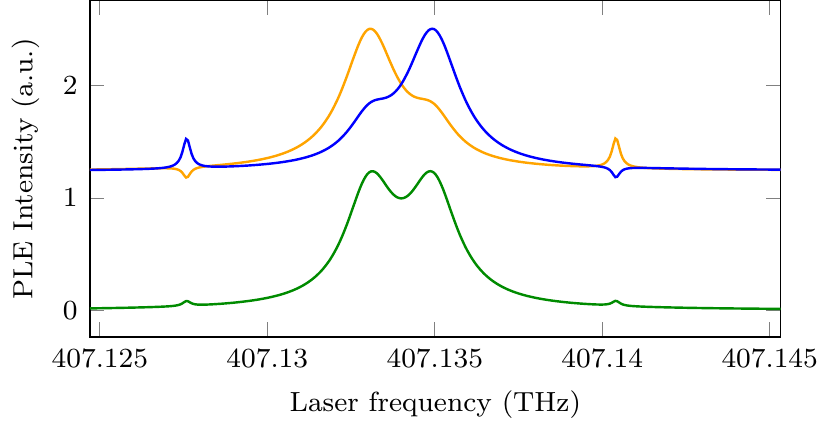}
\caption{
	Simulated spectra for line D showing qualitative correspondence to the measured spectra in \autoref{fig:spectra}(d).
	The rates used in this simulation are shown in \autoref{fig:rates}.
}
\label{fig:simu}
\end{figure}

The equilibrium population distribution for a given laser frequency was obtained by solving the set of differential equations connecting the levels via the rates.
The expected steady-state fluorescence was extracted by summing the population decay from the excited states.
By performing this calculation for a variety of laser frequencies it was possible to obtain simulated excitation spectra as shown in \autoref{fig:simu}.
These show qualitative agreement with the measured excitation spectra, confirming that all of the spectral features can be explained by population dynamics.

\subsubsection*{Spin dynamics in misaligned field}

\begin{figure}[t]
\includegraphics[width=\columnwidth]{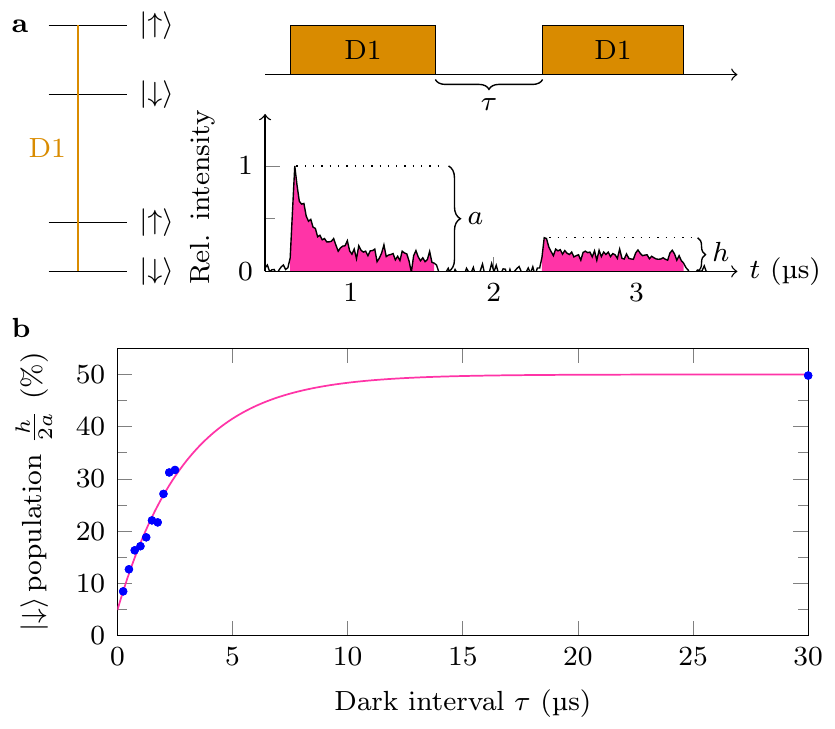}
\caption{
	Optical initialization and relaxation of $\SiV$ spin for a magnetic field misaligned by about 20 degrees.
	(a) A laser pulse resonant to transition D1 provides some fluorescence before flipping the spin state.
	The height $a$ of this first peak is the asymptotic maximum achieved after a long time in the dark.
	Subsequent pulses after short dark intervals $\tau$ had reduced leading-edge height $h$.
	(b) The recovery of this height $h$ with increasing $\tau$ gives the spin relaxation time to be $\ToneSpin=\SI{3.4}{\micro\second}$ in this field configuration.
	Interpreting the asymptotic limit $a$ to correspond to 50\% (thermal) population of the $\ket{\downarrow}$ spin state indicates a spin initialization fidelity of 95\%.
}
\label{fig:high_initialisation}
\end{figure}

Time-resolved fluorescence was used to measure the spin relaxation time with the magnetic field misaligned by about 20 degrees.
This process was equivalent to the measurements reported in \autoref{fig:relaxation_rates}.
The increased off-axis field in this configuration gave the spin-flipping transition D1 a stronger transition dipole and therefore enough cyclicity to produce measureable photons before pumping to the steady-state dark level.
This is seen for the first pulse shown in \autoref{fig:high_initialisation}(a).
The subsequent decay of this fluorescence over the duration of a laser pulse exciting D1 is interpreted as optical pumping to the $\ket{\uparrow}$ spin state.

When a second excitation pulse is applied to transition D1 after some dark interval $\tau$ the height of the leading edge $h$ indicates the amount of relaxation back into the $\ket{\downarrow}$ spin state.
The recovery of the leading edge height $h$ with increasing pulse separation $\tau$ establishes the spin relaxation time to be $\ToneSpin=\SI{3.4}{\micro\second}$ for this field misalignment.
Significantly, for short $\tau$ the leading edge height $h$ becomes the same as the residual steady-state signal, and indicates only 5\% spin population in the $\ket{\downarrow}$ state.
Since the relaxation time of this pumping is far longer than the orbital relaxation time, it is concluded that the laser excitation of D1 has led to spin pumping into the $\ket{\uparrow}$ state with an initalization fidelity of 95\%.

\subsubsection*{CPT on an orbital $\Lambda$-scheme}

\begin{figure}[]
\includegraphics[width=\columnwidth]{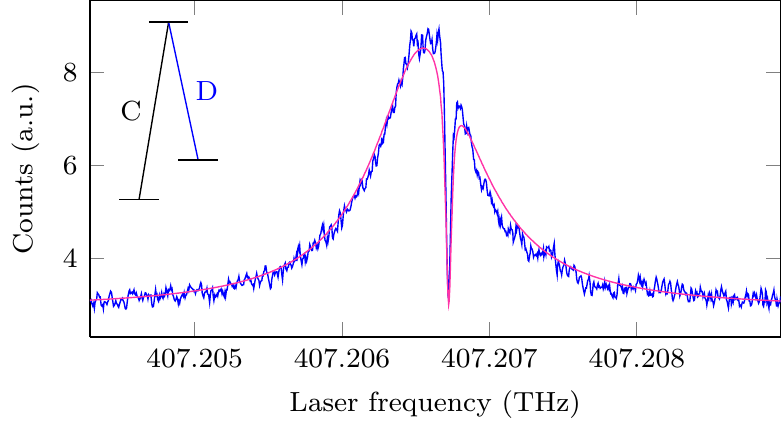}
\caption{
	CPT dip observed for an orbital $\Lambda$-scheme.
	This was measured in the absence of any magnetic field, and so the spin degeneracies were not lifted.
	A pump laser was applied to transition C while a probe laser scanned across transition D.
}
\label{fig:orb_eit}
\end{figure}

All of the CPT features reported in this manuscript for spin levels have required the presence of a magnetic field to achieve Zeeman splitting.
We also observed CPT on the simpler $\Lambda$-scheme comprised of transitions C and D in the absence of any magnetic field.
Transition C was pumped while a second probe laser scanned through the resonance of transition D, and a sharp dip with high contrast was observed as shown in \autoref{fig:orb_eit}.
The clarity of this CPT dip is intriguing, given the short orbital relaxation times established in this manuscript.

\end{document}